\documentclass{ws-procs9x6-cpt22}
\begin{document}

\newcommand{\refeq}[1]{(\ref{#1})}
\def\etal {{\it et al.}}

\title{Lorentz-symmetry violation in scenarios of non-linear electromagnetic models: a preliminary inspection}

\author{P. Gaete,$^1$ and J. A. Helay\"el-Neto$^2$}

\address{$^1$Departamento de F\'{i}sica and Centro Cient\'{i}fico-Tecnol\'ogico de Valpara\'{i}so-CCTVal,\\
Universidad T\'{e}cnica Federico Santa Mar\'{i}a, Valpara\'{i}so, Chile}

\address{$^2$Centro Brasileiro de Pesquisas F\'isicas,\\
Rua Dr. Xavier Sigaud
150, Urca, Rio de Janeiro, Brasil, CEP 22290-180}

\begin{abstract}
In this contribution, our efforts consist in presenting and discussing the status of a paper in progress we are working on to investigate how non-linear electromagnetic effects couple to the parameters that signal Lorentz-symmetry violation (LSV). Here, we realize LSV by means of a specific model, namely Carroll-Field-Jackiw's. We set the formulation by considering a general non-linear photonic Lagrangian (written in terms of the Lorentz-invariant bilinears in the field-strength) that may be coupled to different operators that carry the message of LSV. In possess of the polynomial equation expressing the dispersion relation and the refractive index, one can find the results that express how the (meta)material constitutive properties of the vacuum are affected by the mixing of the parameters that measure LSV with the parameters of the specific non-linear electrodynamic model under consideration.  We expect that our efforts might be of interest in connection with the experiments carried out with the existing super-intense LASERs based on chirped pulse amplification.
\end{abstract}

\bodymatter

\section{Introduction}

As is well known, Lorentz invariance is a fundamental ingredient of quantum field theory, which is an exact symmetry carried out by the Standard Model (SM) \cite{Kostelecky} of the interactions among the smallest building blocks of matter. This theory relies on Lorentz symmetry and provides a remarkably successful description of nowadays known phenomena. Interestingly enough, despite its experimental success, the observation of Lorentz symmetry violation (LSV) would indicate the existence of a new physics. In this form, the possibility that Lorentz and CPT symmetries be spontaneously broken at very fundamental scale, such as in string theories, has motivated a very intensive research activity. More precisely, the necessity of a new scenario has been proposed to overcome theoretical difficulties in the quantum gravity framework \cite{Amelino, Alfaro,Piran,Liberati}. It is worth recalling that theories with LSV are to be considered as effective theories and the analysis of its physical consequences at low energies may provide information and impose constraints on a more fundamental theory. It is to be specially noted that a suitable framework for testing the low-energy manifestations of LSV is the effective approach referred to as the Standard Model Extension (SME), where it is possible to realize spontaneous LSV.

On the other hand, quantum vacuum nonlinearities and its physical consequences such as vacuum birefringence and vacuum dichroism have been of great interest since its earliest days \cite{Euler,Adler,Tollis,Biswas, Seco, Michinel}. The work due to Euler and Heisenberg \cite{Euler} is the key example, who computed an effective nonlinear electromagnetic theory in vacuum emerging from the interaction of photons with virtual electron-positron pairs. Nevertheless, this amazing quantum characteristics of light has generated a growing interest on the experimental side \cite{Bamber,Burke,Pike}. For example, the PVLAS collaboration \cite{Valle}, and more recently the ATLAS collaboration has reported on the direct detection of the light-by-light scattering in LHC Pb-Pb collisions \cite{Aaboud, Enterria}. The advent of laser facilities has given rise to various proposals to probe quantum vacuum nonlinearities \cite{Battesti,Ataman}.

We would like to point out that the work of Ref. \cite{Ferrari} sets out an interesting discussion that relates LSV and non-linear electromagnetic effects. In their paper, LSV is present at the tree-level action in a term that non-minimally couples the photon to charged fermions. The latter are integrated out and thereby an effective action is attained with higher derivative operators. Our present case calls into question a different path: a non-linear action is present from the very onset and a LSV term (actually, CFJ) as well. We then work out the photonic dispersion relations to get how the parameters associated to the non-linearity interfere with the CFJ external vector in the expressions for the refractive indices and the group velocity. The fourth-degree polynomial in the frequency can be shown to split into four categories: a purely Maxwellian term, a piece exclusively given by the non-linearity, a third part isolating the CFJ parameter and, finally, the desired contribution that couples LSV with the CFJ external vector. From this fourth-degree equation in the frequency, one can attain the expressions for the refractive indices and group velocity, from which we are able to read off the explicit form of the terms that couple the non-linearity to the LSV parameters. 

Also, from the preceding considerations, and given the ongoing experiments related to photon-photon interaction physics, we are encouraged to further examine how nonlinear electromagnetic effects couple to the parameters that signal LSV. Specifically, we are concerned with birefringence, as well as the static potential along the lines \cite{inverse,GaeteHel}.

\section{Model under consideration}

We start off our considerations with a brief description of the model under consideration.The model is described by the following Lagrangian density: 
\begin{equation}
{\cal L} = {{\cal L}_{NL}}\left( {{\cal F};{\cal G}} \right) + {{\cal L}_{CFJ}},
 \label{NLCFJ05}
 \end{equation}
 where ${\cal F} \equiv  - \frac{1}{4}{F_{\mu \nu }}{F^{\mu \nu }} = \frac{1}{2}\left( {\frac{{{{\bf E}^2}}}{{{c^2}}} - {{\bf B}^2}} \right)$, ${\cal G} \equiv  - \frac{1}{4}{F_{\mu \nu }}{\tilde F^{\mu \nu }} = \frac{{\bf E}}{c} \cdot {\bf B}$ and ${{\cal L}_{NL}}$ describes the nonlinear part, whereas the term ${{\cal L}_{CFJ}}$ (CFJ denotes Carroll, Field, Jackiw), is given by
 \begin{equation}
{\cal L}_{CFJ} = \frac{1}{4}{\varepsilon ^{\mu \nu \kappa \lambda }}{v_\mu }{A_\nu }{F_{\kappa \lambda }}.  \label{NLCFJ10}
 \end{equation}
 
Next, after splitting ${{\cal L}_{NL}}\left( {{\cal F};{\cal G}} \right)$ around a constant and homogeneous electromagnetic background, ${{\bf E}_B}( {{\bf B}_B} )$, and a small fluctuation
${\bf e}( {\bf b})$,
 (${\bf E} = {{\bf E}_B} + {\bf e}, {\bf B} = {{\bf B}_B} + {\bf b}$), the corresponding equations of motion up to quadratic terms read 
\begin{eqnarray}
\nabla  \cdot {\bf d} - {\bf v} \cdot {\bf b} = 0, \nonumber\\
\nabla  \times {\bf e} =  - \frac{{\partial {\bf b}}}{{\partial t}}, \nonumber\\
\nabla  \cdot {\bf b} = 0, \nonumber\\
\nabla  \times {\bf h} - {v^0}\, {\bf b} + {\bf v} \times {\bf e} = \frac{{\partial {\bf e}}}{{\partial t}},  \label{NLCFJ15}
\end{eqnarray}  
where the ${\bf d}$ and ${\bf h}$ fields are given by
\begin{equation}
 {\bf d} = \mathord{\buildrel{\lower3pt\hbox{$\scriptscriptstyle\leftrightarrow$}} 
\over \varepsilon }  \cdot {\bf e} + \mathord{\buildrel{\lower3pt\hbox{$\scriptscriptstyle\leftrightarrow$}} 
\over \zeta }  \cdot {\bf b}, \label{NLCFJ20}
\end{equation}
\begin{equation}
{\bf h} = {\mathord{\buildrel{\lower3pt\hbox{$\scriptscriptstyle\leftrightarrow$}} 
\over \mu } ^{ - 1}} \cdot {\bf b} + \mathord{\buildrel{\lower3pt\hbox{$\scriptscriptstyle\leftrightarrow$}} 
\over \eta }  \cdot {\bf e}. \label{NLCFJ25}
\end{equation}
It should be emphasized again that the vacuum electromagnetic properties are characterized by the following expressions for the vacuum permittivity and the vacuum permeability:
\begin{equation}
{\varepsilon _{ij}} \equiv {\delta _{ij}} + {\alpha _i}{E_j} + {\beta _i}{B_j},  \label{NLCFJ30}
\end{equation}
\begin{equation}
{\mu ^{ - 1}}_{ij} \equiv {\delta _{ij}} - {B_i}{\gamma _j} - {E_i}{\Delta _j}.  \label{NLCFJ35}
\end{equation}
Here we have simplified our notation by writing
\begin{equation}
\boldsymbol {\alpha}  \equiv \frac{1}{{{C_1}}}\left( {{D_1}\,{\bf E} + {D_3}\,{\bf B}} \right), \ \ \ \boldsymbol {\beta}  \equiv \frac{1}{{{C_1}}}\left( {{D_2}\, {\bf B} + {D_3} \,{\bf E}} \right),  \label{NLCFJ40}
\end{equation}
\begin{equation}
\boldsymbol {\gamma}  \equiv \frac{1}{{{C_1}}}\left( {{D_1}\, {\bf B} - {D_3} \,{\bf E}} \right), \ \ \ \boldsymbol {\Delta}  \equiv \frac{1}{{{C_1}}}\left( { - {D_3}\, {\bf B} + {D_2} \,{\bf E}} \right), \label{NLCFJ45}
\end{equation}
where
\begin{equation}
{C_1} \equiv \frac{{\partial {{\cal L}_{NL}}}}{{\partial {\cal F}}}, \ \ \ {D_1} \equiv \frac{{{\partial ^2}{{\cal L}_{NL}}}}{{\partial {{\cal F}^2}}}, \ \ \ {D_2} \equiv \frac{{{\partial ^2}{{\cal L}_{NL}}}}{{\partial {{\cal G}^2}}}, \ \ {D_3} \equiv \frac{{{\partial ^2}{{\cal L}_{NL}}}}{{\partial {\cal F}\partial {\cal G}}}. \label{NLCFJ50}
\end{equation}
 
Mention should be made, at this point, to a further aspect related to vacuum electromagnetic properties. More specifically, we refer to:
 How do the parameters and external fields coming from the nonlinear (NL) sector couple to LSV parameters ($v^{\mu}$ for CFJ)? In what follows we will examine this question.

\section{Dispersion relations} 

In order to adequately deal with this issue, one must inspect the dispersion relations (DRs) for an electromagnetic wave propagating in the external electromagnetic background; for the model under consideration, the matrix yielding the DRs reads as follows below:
\begin{equation}
\det {M_{ij}}\left( {w,{\bf k}; {\bf E}, {\bf B};{v^\mu }} \right) = 0, \label{NLCFJ55}
\end{equation}
where the $M$-matrix has the form
\begin{eqnarray}
{M_{ij}} &\equiv& {w^2}{\varepsilon _{ij}} + w{\zeta _{im}}{\varepsilon _{mnj}}{k_n} + {\varepsilon _{imn}}{\varepsilon _{klj}}{\mu ^{ - 1}}_{nk}{k_m}{k_l} \nonumber\\
 &+& w{\varepsilon _{imn}}{\eta _{nj}}{k_m} + i{v^0}{\varepsilon _{imj}}{k_m} - iw{\varepsilon _{imj}}{v_m}. \label{NLCFJ60}
 \end{eqnarray}

Next, we shall focus here on a single situation for the different nonlinear (NL) models:
${\bf E_{B}}= 0, {\bf B_{B}} \ne 0$. This decouples the coefficient $D_{3}$ from the problem and the $M$-matrix becomes:
\begin{eqnarray}
{M_{ij}} &=& \left[ {\left( {{w^2} - {{\bf k}^2}} \right) + \frac{{{D_1}}}{{{C_1}}}\left( {{\bf B}_B^2\,
{{\bf k}^2} - {{\left( {{{\bf B}_B} \cdot {\bf k}} \right)}^2}} \right)} \right]{\delta _{ij}} \nonumber\\
 &+&\left( {\frac{{{D_2}}}{{{C_1}}}{w^2} - \frac{{{D_1}}}{{{C_1}}}{{\bf k}^2}} \right){B_{Bi}}{B_{Bj}} + \left( {1 - \frac{{{D_1}}}{{{C_1}}}{\bf B}_B^2} \right){k_i}{k_j} \nonumber\\
 &+& \frac{{{D_1}}}{{{C_1}}}\left( {{\bf B}_{B} \cdot {\bf k}} \right)\left( {{k_i}{B_{Bj}} + {k_j}{B_{Bi}}} \right) \nonumber\\
&+& i{\varepsilon _{imj}}\left( {{v^0}{k_m} - w{v_m}} \right).  \label{NLCFJ65}
\end{eqnarray} 

Interestingly, the matrix from which the dispersion relations (DRs) follow can then be split into $4$ pieces:
\begin{equation}
DR= Max +NL+LS+ NL/LSV=0,   \label{NLCFJ70}
\end{equation}
where Max, NL, LSV and NL/LSV denote Maxwell, Nonlinear, Lorentz symmetry violation and the coupling between nonlinear and LSV, respectively.
These terms are given by:
\begin{equation}
Max = \left( {{w^2} - {{\bf k}^2}} \right)^{2},  \label{NLCFJ70a}
\end{equation}
\begin{eqnarray}
NL &=& \frac{{{D_2}}}{{{C_1}}}{{\bf B}_{B}^2}\,{w^4} - \frac{{{D_1}}}{{{C_1}}}{{\bf B}_{B}^2}\,{{\bf k}^4} \nonumber\\
&+& \left( {\frac{{{D_1}}}{{{C_1}}} - \frac{{{D_2}}}{{{C_1}}} + \frac{{{D_1}{D_2}}}{{C_1^2}}{{\bf B}_{B}^2}} \right){{\bf B}_{B}^2}\,{{\bf k}^2}\,{w^2},  \label{NLCFJ70b}
\end{eqnarray}
\begin{equation}
LSV =  - {\bf v^2}{w^2} + {{\bf v}^2}{{\bf k}^2} - {\left( {{\bf v} \cdot {\bf k}} \right)^2}. \label{NLCFJ70c}
\end{equation}

The coupled NL/LSV effects we are pursuing are cast in what follows:
\begin{eqnarray}
NL/LSV &\equiv&  - \frac{{{D_2}}}{{{C_1}}}{\left( {{{\bf B}_B} \cdot {\bf v}} \right)^2}{w^2} + \frac{{{D_1}}}{{{C_1}}}{\left( {{{\bf B}_B} \cdot {\bf v}} \right)^2}{\bf k^2} \nonumber\\
&-&\frac{{{D_1}}}{{{C_1}}}{\bf B}^2_{B}\left( {{{\bf v}^2}{{\bf k}^2} - {{\left( {{\bf v} \cdot {\bf k}} \right)}^2}} \right). \label{NLCFJ75}
\end{eqnarray}

It is worthy recalling, at this stage, that the coefficients $C_{1}$, $D_{1}$, $D_{2}$ specify the particular nonlinear electrodynamic model under consideration; where ${C_1}=\frac{{\partial {{\cal L}_{NL}}}}{{\partial {\cal F}}}$, ${D_1}=\frac{{{\partial ^2}{{\cal L}_{NL}}}}{{\partial {{\cal F}^2}}}$, ${D_2} = \frac{{{\partial ^2}{{\cal L}_{NL}}}}{{\partial {{\cal G}^2}}}$ and $D_{3}$ does not contribute whenever either ${\bf E}_{B}$ or ${\bf B}_{B}$ vanishes.

In this way, one encounters
\begin{equation}
m_\gamma ^2 = \frac{{{{\bf v}^2} - \frac{{{D_2}}}{{{C_1}}}{{\left( {{{\bf B}_B} \cdot {\bf v}} \right)}^2}}}{{1 + \frac{{{D_2}}}{{{C_1}}}{\bf B}_B^2}}.  \label{NLCFJ80}
\end{equation}
which represents, in the particle picture of the propagating mode, the photon (inertial) mass, which now depends on both sets of non-linear and LSV parameters and also on the direction of the external magnetic field relative to the external CFJ space vector.
Proceeding further, the group velocity can be shown to result from the composition between the wave vector and the external CFJ space vector as written in the expression below:
\begin{equation}
{{\bf v}_g} = \frac{{\bf k}}{w}\,\frac{P}{Q} + \frac{{\bf v}}{w}\,\frac{R}{Q}.  \label{NLCFJ85}
\end{equation}
To get the last line we used
\begin{eqnarray}
P &\equiv&{w^2} - {{\bf k}^2} - \frac{1}{2}\left( {\frac{{{D_1}}}{{{C_1}}} - \frac{{{D_2}}}{{{C_1}}} + \frac{{{D_1}{D_2}}}{{C_1^2}}} \right){\bf B}_B^2\,{w^2} \nonumber\\
 &+& \frac{{{D_1}}}{{{C_1}}}{\bf B}_B^2\,{{\bf k}^2} - \frac{1}{2}{{\bf v}^2} + \frac{1}{2}\frac{{{D_1}}}{{{C_1}}}{\left( {{{\bf B}_B} \cdot {\bf v}} \right)^2} - \frac{1}{2}\frac{{{D_1}}}{{{C_1}}}{\bf B}_B^2\,{{\bf v}^2}, \label{NLCFJ90}
\end{eqnarray}
\begin{eqnarray}
Q &\equiv& {w^2} - {{\bf k}^2} + \frac{1}{2}\left( {\frac{{{D_1}}}{{{C_1}}} - \frac{{{D_2}}}{{{C_1}}} + \frac{{{D_1}{D_2}}}{{C_1^2}}{\bf B}_B^2} \right){\bf B}_B^2\,{{\bf k}^2} \nonumber\\
&+&\frac{{{D_2}}}{{{C_1}}}{\bf B}_B^2\,{w^2} - \frac{1}{2}{{\bf v}^2} - \frac{1}{2}\frac{{{D_2}}}{{{C_1}}}{\left( {{{\bf B}_B} \cdot {\bf v}} \right)^2},  \label{NLCFJ95}
\end{eqnarray}
\begin{equation}
R \equiv \frac{1}{2}{\bf v} \cdot {\bf k} - \frac{1}{2}\frac{{{D_1}}}{{{C_1}}}{\bf B}_B^2\left( {{\bf v} \cdot {\bf k}} \right).  \label{NLCFJ100}
\end{equation}

The refractive indices, ${n_\bot}$ and ${n_\parallel}$, can also be found with the help of our DRs, given by det M = 0. This is not presented here, but we shall work them out for the Born-Infeld, Euler-Heisenberg, Hoffmann-Infeld, Logarithmic and the ModMax models in the extended version of this contribution, to be submitted elsewhere in a near future.

\section{Final remarks}

In summary, in this presentation we have reported on the results that take into account how the (meta)material constitutive properties of the vacuum are affected by the parameters that measure LSV.
 Finally, we have calculated interaction potential for logarithmic electrodynamics with a term CFJ. Our calculations show that the interaction energy, at leading order in $\beta ^{2}$, is similar to our previous calculation \cite{GaeteHel}:
\begin{equation}
V =  - \frac{{{q^2}}}{{4\pi }}\frac{1}{\eta }\frac{1}{r} + \frac{{{q^2}}}{{4\pi }}\frac{{{\mu ^2}{{\bf v}^2}}}{{16\eta }}z\ln \left( {\frac{{z + r}}{{2z}}} \right),  \label{NLCFJ105}
\end{equation}
where $r = \sqrt {{{\bf x}^2} + {{\bf y}^2}}$. In this case, ${\bf E_{B}}=0$, ${\bf B_{B}} = B {\hat {\bf z}}$ and ${\bf v} = v {\hat {\bf z}}$. Whereas $\mu$ and $\frac{1}{\eta }$ are given by
\begin{equation}
\mu  = \sqrt {\frac{{1 + {\raise0.7ex\hbox{${{{\bf B}^2}}$} \!\mathord{\left/
 {\vphantom {{{{\bf B}^2}} {{\beta ^2}}}}\right.\kern-\nulldelimiterspace}
\!\lower0.7ex\hbox{${{\beta ^2}}$}}}}{{1 - {\raise0.7ex\hbox{${{{\bf B}^2}}$} \!\mathord{\left/
 {\vphantom {{{{\bf B}^2}} {{\beta ^2}}}}\right.\kern-\nulldelimiterspace}
\!\lower0.7ex\hbox{${{\beta ^2}}$}}}}} \left( {1 + {\raise0.7ex\hbox{${{{\bf B}^2}}$} \!\mathord{\left/
 {\vphantom {{{{\bf B}^2}} {2{\beta ^2}}}}\right.\kern-\nulldelimiterspace}
\!\lower0.7ex\hbox{${2{\beta ^2}}$}}} \right), \label{NLCFJ110}
\end{equation}
\begin{equation}
\frac{1}{\eta } = \frac{{\left( {1 + {\raise0.7ex\hbox{${{{\bf B}^2}}$} \!\mathord{\left/
 {\vphantom {{{{\bf B}^2}} {2{\beta ^2}}}}\right.\kern-\nulldelimiterspace}
\!\lower0.7ex\hbox{${2{\beta ^2}}$}}} \right)}}{{\left( {1 - {\raise0.7ex\hbox{${{{\bf B}^2}}$} \!\mathord{\left/
 {\vphantom {{{{\bf B}^2}} {{\beta ^2}}}}\right.\kern-\nulldelimiterspace}
\!\lower0.7ex\hbox{${{\beta ^2}}$}}} \right)}}. \label{NLCFJ115}
\end{equation}
In an extended paper that we shall submit elsewhere \cite{GaeteHel2}, we are pursuing the investigation of other LSV photonic operators  (here, we have restricted ourselves to the CFJ case) to inspect how they couple to the parameters of the non-linear electromagnetic sector in the presence of an electric/magnetic background to estimate how a very strong background may enhance the tiny interference terms. We expect that our efforts in this endeavor might allow us to get bounds on the model parameters by consulting results from the experiments carried out with the existing super-intense LASERs based on chirped pulse amplification. 

\section*{Acknowledgments}
One of us (P.G.) was partially supported by ANID PIA / APOYO AFB180002.

\end{document}